\documentclass[twocolumn]{article}
\usepackage{epsfig}
\newcommand{\be}{\begin{equation}}
\newcommand{\ee}{\end{equation}}
\newcommand{\bea}{\begin{eqnarray}}
\newcommand{\eea}{\end{eqnarray}}
\newcommand{\ba}{\begin{array}}
\newcommand{\ea}{\end{array}}

\begin{document}
\title{Hierarchical Clustering Based on Mutual Information}
\author{Alexander Kraskov, Harald St\"ogbauer, Ralph G. Andrzejak, and Peter Grassberger \\
John-von-Neumann Institute for Computing, \\ Forschungszentrum J\"ulich,D-52425 J\"ulich, Germany}

\date{\today}
\maketitle
\begin{abstract}
\noindent \textbf{Motivation:} Clustering is a frequently used concept in variety of bioinformatical
applications. We present a new method for hierarchical clustering of data called {\it mutual information
clustering} (MIC) algorithm. It uses mutual information (MI) as a similarity measure and exploits its grouping
property: The MI between three objects $X, Y,$ and $Z$ is equal to the sum of the MI between $X$ and $Y$, plus
the MI between $Z$ and the combined object $(XY)$.

\noindent \textbf{Results:} We use this both in the Shannon (probabilistic) version of information theory, where
the ``objects" are probability distributions represented by random samples, and in the Kolmogorov (algorithmic)
version, where the ``objects" are symbol sequences. We apply our method to the construction of mammal
phylogenetic trees from mitochondrial DNA sequences and we reconstruct the fetal ECG from the output of
independent components analysis (ICA) applied to the ECG of a pregnant woman.

\noindent \textbf{Availability:} The programs for estimation of MI and for clustering (probabilistic version)
are available at \textsf{http://www.fz-juelich.de/nic/cs/software}.

\noindent \textbf{Contact:} \textsf{a.kraskov@fz-juelich.de}
\end{abstract}

%
%
%

\section{Introduction}

Classification or organizing of data is very important in all scientific disciplines. It is one of the most
fundamental mechanism of understanding and learning \cite{jain-dubes}. Depending on the problem, classification
can be exclusive or overlapping, supervised or unsupervised. In the following we will be interested only in
exclusive unsupervised classification. This type of classification is usually called clustering or cluster
analysis.

An instance of a clustering problem consist of a set of objects and a set of properties (called characteristic
vector) for each object. The goal of clustering is the separation of objects into groups using only the
characteristic vectors. Indeed, in general only certain aspects of the characteristic vectors will be relevant,
and extracting these relevant features is one field where mutual information (MI) plays a major role
\cite{bottleneck}, but we shall not deal with this here. Cluster analysis organizes data either as a single
grouping of individuals into non-overlapping clusters or as a hierarchy of nested partitions. The first approach
is called partitional clustering (PC), the second one is hierarchical clustering (HC). One of the main features
of HC methods is the visual impact of the {\it dendrogram} which enables one to see how objects are being merged
into clusters. From any HC one can obtain a PC by restricting oneself to a ``horizontal" cut through the
dendrogram, while one cannot go in the other direction and obtain a full hierarchy from a single PC. Because of
their wide spread of applications, there are a large variety of different clustering methods in use
\cite{jain-dubes}.

The crucial point of all clustering algorithms is the choice of a {\it proximity measure}. This is obtained from
the characteristic vectors and can be either an indicator for similarity (i.e. large for similar and small for
dissimilar objects), or dissimilarity. In the latter case it is convenient but not obligatory if it satisfies
the standard axioms of a metric (positivity, symmetry, and triangle inequality). A matrix of all pairwise
proximities is called proximity matrix. Among HC methods one should distinguish between those where one uses the
characteristic vectors only at the first level of the hierarchy and derives the proximities between clusters
from the proximities of their constituents, and methods where the proximities are calculated each time from
their characteristic vectors. The latter strategy (which is used also in the present paper) allows of course for
more flexibility but might also be computationally more costly.

Quite generally, the ``objects" to be clustered can be either single (finite) patterns (e.g. DNA sequences) or
random variables, i.e. {\it probability distributions}. In the latter case the data are usually supplied in form
of a statistical sample, and one of the simplest and most widely used similarity measures is the linear
(Pearson) correlation coefficient. But this is not sensitive to nonlinear dependencies which do not manifest
themselves in the covariance and can thus miss important features. This is in contrast to mutual information
(MI) which is also singled out by its information theoretic background \cite{cover-thomas}. Indeed, MI is zero
only if the two random variables are strictly independent.

Another important feature of MI is that it has also an ``algorithmic" cousin, defined within algorithmic
(Kolmogorov) information theory \cite{li-vi} which measures the similarity between individual objects. For a
thorough discussion of distance measures based on algorithmic MI and for their application to clustering, see
\cite{li1,li2}.

Another feature of MI which is essential for the present application is its {\it grouping property}: The MI
between three objects (distributions) $X, Y,$ and $Z$ is equal to the sum of the MI between $X$ and $Y$, plus
the MI between $Z$ and the combined object (joint distribution) $(XY)$,
\be
   I(X,Y,Z) = I(X,Y) + I((X,Y),Z).                        \label{group}
\ee
Within Shannon information theory this is an exact theorem (see below), while it is true in the algorithmic
version up to the usual logarithmic correction terms \cite{li-vi}. Since $X,Y,$ and $Z$ can be themselves
composite, Eq.(\ref{group}) can be used recursively for a cluster decomposition of MI. This motivates the main
idea of our clustering method: instead of using e.g. centers of masses in order to treat clusters like
individual objects in an approximative way, we treat them exactly like  individual objects when using MI as
proximity measure.

More precisely, we propose the following scheme for clustering $n$ objects with MIC:\\
(1) Compute a proximity matrix based on pairwise mutual informations; assign $n$ clusters such that each cluster
contains exactly one object;\\
(2) find the two closest clusters $i$ and $j$; \\
(3) create a new cluster $(ij)$ by combining $i$ and $j$; \\
(4) delete the lines/columns with indices $i$ and $j$ from the proximity matrix, and add one line/column
containing the proximities between cluster $(ij)$ and all
other clusters; \\
(5) if the number of clusters is still $>2$, goto (2); else join the two clusters and stop.

In the next section we shall review the pertinent properties of MI, both in the Shannon and in the algorithmic
version. This is applied in Sec.~3 to construct a phylogenetic tree using mitochondrial DNA and in Sec.~4 to
cluster the output channels of an independent component analysis (ICA) of an electrocardiogram (ECG) of a
pregnant woman, and to reconstruct from this the maternal and fetal ECGs. We finish with our conclusions in
Sec.~6.

\section{Mutual Information}

\subsection{Shannon Theory}

Assume that one has two random variables $X$ and $Y$. If they are discrete, we write $p_i(X) = {\rm
prob}(X=x_i)$, $p_i(Y) = {\rm prob}(Y=x_i)$, and $p_{ij} = {\rm prob}(X=x_i,Y=y_i)$ for the marginal and joint
distribution. Otherwise (and if they have finite densities) we denote the densities by $\mu_X(x),\mu_Y(y)$, and
$\mu(x,y)$. Entropies are defined for the discrete case as usual by $H(X) = - \sum_ip_i(X) \log p_i(X)$, $H(Y) =
- \sum_ip_i(Y) \log p_i(Y)$, and $H(X,Y)=-\sum_{i,j} p_{ij} \log p_{ij}$. Conditional entropies are defined as
$H(X|Y) = H(X,Y)-H(Y) = -\sum_{i,j} p_{ij} \log p_{i|j}$. The base of the logarithm determines the units in
which information is measured. In particular, taking base two leads to information measured in bits. In the
following, we always will use natural logarithms. The MI between $X$ and $Y$ is finally defined as
\bea
   I(X,Y) &=& H(X)+H(Y)-H(X,Y) \nonumber  \\
          &=& \sum_{i,j} p_{ij}\;\log{p_{ij}\over p_i(X)p_i(Y)}.
\eea
It can be shown to be non-negative, and is zero only when $X$ and $Y$ are strictly independent. For $n$ random
variables $X_1,X_2\ldots X_n$, the MI is defined as
\be
   I(X_1,\ldots, X_n) = \sum_{k=1}^n H(X_k) - H(X_1,\ldots, X_n).
\ee This quantity is often referred to as (generalized) redundancy, in order to distinguish it from different
``mutual informations" which are constructed analogously to higher order cumulants, but we shall not follow this
usage. Eq.(\ref{group}) can be checked easily, together with its generalization to arbitrary groupings. It means
that MI can be {\it decomposed into hierarchical levels}. By iterating it, one can decompose $I(X_1\ldots X_n)$
for any $n>2$ and for any partitioning of the set $(X_1\ldots X_n)$ into the MIs between elements within one
cluster and MIs between clusters.

For continuous variables one first introduces some binning (`coarse-graining'), and applies the above to the
binned variables. If $x$ is a vector with dimension $m$ and each bin has Lebesgue measure $\Delta$, then $p_i(X)
\approx \mu_X(x)\Delta^m$ with $x$ chosen suitably in bin $i$, and \footnote{Notice that we have here assumed
that densities really exists. If not e.g. if $X$ lives on a fractal set), then $m$ is to be replaced by the
Hausdorff dimension of the measure $\mu$.}
\be
   H_{\rm bin}(X) \approx \tilde{H}(X) - m \log \Delta
\ee
where the {\it differential entropy} is given by
\be
   \tilde{H}(X) = -\int dx \;\mu_X(x) \log \mu_X(x).
\ee
Notice that $H_{\rm bin}(X)$ is a true (average) information and is thus non-negative, but $\tilde{H}(X)$ is not
an information and can be negative. Also, $\tilde{H}(X)$ is not invariant under homeomorphisms $x\to \phi(x)$.

Joint entropies, conditional entropies, and MI are defined as above, with sums replaced by integrals. Like
$\tilde{H}(X)$, joint and conditional entropies are neither positive (semi-)definite nor invariant. But MI,
defined as
\be
   I(X,Y) = \int\!\!\!\int dx dy \;\mu_{XY}(x,y) \;\log{\mu_{XY}(x,y)\over \mu_X(x)\mu_Y(y)}\;,
   \label{mi}
\ee
is non-negative and invariant under $x\to \phi(x)$ and $y\to \psi(y)$. It is (the limit of) a true information,
\be
   I(X,Y) = \lim_{\Delta\to 0} [H_{\rm bin}(X)+H_{\rm bin}(Y)-H_{\rm bin}(X,Y)].
\ee

In applications, one usually has the data available in form of a statistical sample. To estimate $I(X,Y)$ one
starts from $N$ bivariate measurements $(x_i,y_i), \, i=1,\ldots N$ which are assumed to be iid (independent
identically distributed) realizations. There exist numerous algorithms to estimate $I(X,Y)$ and entropies. We
shall use in the following the MI estimators proposed recently in Ref.~\cite{mi}, and we refer to this paper for
a review of alternative methods.

\subsection{Algorithmic Information Theory}

In contrast to Shannon theory where the basic objects are random variables and entropies are {\it average}
informations, algorithmic information theory deals with individual symbol strings and with the actual
information needed to specify them. To ``specify" a sequence $X$ means here to give the necessary input to a
universal computer $U$, such that $U$ prints $X$ on its output and stops. The analogon to entropy, called here
usually the {\it complexity} $K(X)$ of $X$, is the minimal length of an input which leads to the output $X$, for
fixed $U$. It depends on $U$, but it can be shown that this dependence is weak and can be neglected in the limit
when $K(X)$ is large \cite{li-vi}.

Let us denote the concatenation of two strings $X$ and $Y$ as $XY$. Its complexity is $K(XY)$. It is intuitively
clear that $K(XY)$ should be larger than $K(X)$ but cannot be larger than the sum $K(X)+K(Y)$. Finally, one
expects that $K(X|Y)$, defined as the minimal length of a program printing $X$ when $Y$ is furnished as
auxiliary input, is related to $K(XY)-K(Y)$. Indeed, one can show \cite{li-vi} (again within correction terms
which become irrelevant asymptotically) that
\be
   0 \leq K(X|Y) \simeq K(XY)-K(Y) \leq K(X).
\ee
Notice the close similarity with Shannon entropy.

The algorithmic information in $Y$ about $X$ is finally defined as
\be
   I_{\rm alg}(X,Y) = K(X) - K(X|Y) \simeq K(X)+K(Y)-K(XY).
\ee
Within the same additive correction terms, one shows that it is symmetric, $I_{\rm alg}(X,Y) = I_{\rm
alg}(Y,X)$, and can thus serve as an analogon to mutual information.

From the halting theorem it follows that $K(X)$ is in general not computable. But one can easily give upper
bounds. Indeed, the length of any input which produces $X$ (e.g. by spelling it out verbatim) is an upper bound.
Improved upper bounds are provided by any file compression algorithm such as gnuzip or UNIX ``compress". Good
compression algorithms will give good approximations to $K(X)$, and algorithms whose performance does not depend
on the input file length (in particular since they do not segment the file during compression) will be crucial
for the following.

\subsection{MI-Based Distance Measures}

Mutual information itself is a similarity measure in the sense that small values imply large ``distances" in a
loose sense. But it would be useful to modify it such that the resulting quantity is a metric in the strict
sense, i.e. satisfies the triangle inequality. Indeed, the first such metric is well known \cite{cover-thomas}:
The quantity
\be
   d(X,Y)=H(X|Y)+H(Y|X)=H(X,Y)-I(X,Y)                   \label{d}
\ee
satisfies the triangle inequality, in addition to being non-negative and symmetric and to satisfying $d(X,X)=0$.
The proof proceeds by first showing that for any $Z$
\be
   H(X|Y) \leq H(X,Z|Y) \leq H(X|Z)+H(Z|Y).                                        \label{lemma0}
\ee

But $d(X,Y)$ is not appropriate for our purposes. Since we want to compare the proximity between two single
objects and that between two clusters containing maybe many objects, we would like the distance measure to be
unbiased by the sizes of the clusters. As argued forcefully in \cite{li1,li2}, this is not true for $I_{\rm
alg}(X,Y)$, and for the same reasons it is not true for $I(X,Y)$ or $d(X,Y)$ either: A mutual information of
thousand bits should be considered as large, if $X$ and $Y$ themselves are just thousand bits long, but it
should be considered as very small, if $X$ and $Y$ would each be huge, say one million bits.

As shown in \cite{li1,li2} within the algorithmic framework, one can form two different distances which measure
{\it relative} distance, i.e. which are normalized by dividing by a total entropy. We sketch here only the
theorems and proofs for the Shannon version, they are indeed very similar to their algorithmic analoga in
\cite{li1,li2}.

\noindent {\sc Theorem 1}: The quantity
\be
   D(X,Y) = 1 - \frac{I(X,Y)}{H(X,Y)} = \frac{d(X,Y)}{H(X,Y)}                 \label{eq:dist}
\ee
is a metric, with $D(X,X)=0$ and $D(X,Y)\leq 1$ for all pairs $(X,Y)$.

\noindent {\sc Proof}: Symmetry, positivity and boundedness are obvious. Since $D(X,Y)$ can be written as
\be
    D(X,Y)=\frac{H(X|Y)}{H(X,Y)}+\frac{H(Y|X)}{H(Y,X)},
\ee
it is sufficient for the proof of the triangle inequality to show that each of the two terms on the r.h.s.  is
bounded by an analogous inequality, i.e.
\be
   \frac{H(X|Y)}{H(X,Y)} \leq \frac{H(X|Z)}{H(X,Z)}+\frac{H(Z|Y)}{H(Z,Y)}   \label{lemma}
\ee
and similarly for the second term. Eq.(\ref{lemma}) is proven straightforwardly, using Eq.(\ref{lemma0}) and the
basic inequalities $H(X) \geq 0$, $H(X,Y) \leq H(X,Y,Z)$ and $H(X|Z)\geq 0$:
\bea
    \frac{H(X|Y)}{H(X,Y)} &=& \frac{H(X|Y)}{H(Y)+H(X|Y)}  \nonumber \\
    & \leq & \frac{H(X|Z)+H(Z|Y)}{H(Y)+H(X|Z)+H(Z|Y)} \nonumber \\
    & = & \frac{H(X|Z)+H(Z|Y)}{H(X|Z)+H(Y,Z)} \nonumber \\
    & \leq &\frac{H(X|Z)}{H(X|Z)+H(Z)}+\frac{H(Z|Y)}{H(Y,Z)} \nonumber \\
    & = & \frac{H(X|Z)}{H(X,Z)}+\frac{H(Z|Y)}{H(Z,Y)}.
\eea

\noindent {\sc Theorem 2}: The quantity
\bea
   D'(X,Y) & = & 1 - \frac{I(X,Y)}{\max\{H(X),H(Y)\}}    \nonumber \\
     & = & \frac{\max\{H(X|Y),H(Y|X)\}}{\max\{H(X),H(Y)\}}                \label{eq:dist2}
\eea
is also a metric, also with $D'(X,X)=0$ and $D'(X,Y)\leq 1$ for all pairs $(X,Y)$. It is sharper than $D$, i.e.
$D'(X,Y) \leq D(X,Y)$.

\noindent
{\sc Proof}: Again we have only to prove the triangle inequality, the other parts are trivial. For this we have to distinguish different cases \cite{li2}.\\
Case 1: $\max\{H(Z),H(Y)\}\leq H(X)$. Using Eq.(\ref{lemma0}) we obtain
\bea
    D'(X,Y) & = & \frac{H(X|Y)}{H(X)} \leq \frac{H(X|Z)}{H(X)} + \frac{H(Z|Y)}{H(Y)}  \nonumber \\
            & = & D'(X,Z)+D'(Z,Y).
\eea
Case 2: $\max\{H(Z),H(X)\}\leq H(Y)$. This is completely analogous.\\
Case 3: $H(X)\leq H(Y)< H(Z)$. We now have to show that
\bea
    D'(X,Y) & = &  \frac{H(Y|X)}{H(Y)} \leq \frac{H(Y|Z)+H(Z|X)}{H(Y)}  \nonumber \\
            & \stackrel{?}{\leq} &  D'(X,Z)+D'(Z,Y) \nonumber \\
            & = &  \frac{H(Z|X)}{H(Z)}+\frac{H(Z|Y)}{H(Z)}.      \label{dd}
\eea
Indeed, if the r.h.s. of the first line is less than 1, then
\bea
    \frac{H(Y|X)}{H(Y)} & \leq &\frac{H(Y|Z)+H(Z|X)}{H(Y)}  \nonumber \\
             &\leq & \frac{H(Y|Z)+H(Z|X)+H(Z)-H(Y)}{H(Z)} \nonumber \\
             & = &   \frac{H(Z|Y)+H(Z|X)}{H(Z)},
\eea
and Eq.(\ref{dd}) holds. If it is larger than 1, then also $(H(Z|Y)+H(Z|X))/H(Z) \geq 1$.
Eq.(\ref{dd}) must now also hold, since $H(Y|X)/H(Y) \leq 1$. \\
Case 4: $H(Y)\leq H(X)< H(Z)$. This is completely analogous to case 3.

Apart from scaling correctly with the total information, in contrast to $d(X,Y)$, the algorithmic analog to
$D(X,Y)$ and $D'(X,Y)$ are also {\it universal} \cite{li2}. Essentially this means that if $X\approx Y$
according to any non-trivial distance measure, then $X\approx Y$ also according to $D$, and even more so (by
factor up to  2) according to $D'$. In contrast to the other properties of $D$ and $D'$, this is not easy to
carry over from algorithmic to Shannon theory. The proof in Ref.~\cite{li2} depends on $X$ and $Y$ being
discrete, which is obviously not true for probability distributions. Based on the universality argument, it was
argued in \cite{li2} that $D'$ should be superior to $D$, but the numerical studies shown in that reference did
not show a clear difference between them. In the following we shall therefore use primarily $D$ for simplicity,
but we checked that using $D'$ did not give systematically better results.

A major difficulty appears in the Shannon framework, if we deal with continuous random variables. As we
mentioned above, Shannon informations are only finite for coarse-grained variables, while they diverge if the
resolution tends to zero. This means that dividing MI by the entropy as in the definitions of $D$ and $D'$
becomes problematic. One has essentially two alternative possibilities. The first is to actually introduce some
coarse-graining, although it would have been necessary for the definition of $I(X,Y)$, and divide by the
coarse-grained entropies. This introduces an arbitrariness, since the scale $\Delta$ is completely ad hoc,
unless it can be fixed by some independent arguments. We have found no such arguments, and thus we propose the
second alternative. There we take $\Delta \to 0$. In this case $H(X) \sim m_x \log \Delta$, with $m_x$ being the
dimension of $X$. In this limit $D$ and $D'$ would tend to 1. But using similarity measures
\bea
   S(X,Y) = (1-D(X,Y))\log(1/\Delta),          \\
   S'(X,Y) = (1-D'(X,Y))\log(1/\Delta)
\eea
instead of $D$ and $D'$ gives {\it exactly} the same results in MIC, and
\be
   S(X,Y) = \frac{I(X,Y)}{m_x+m_y}, \quad S'(X,Y) = \frac{I(X,Y)}{\max\{m_x,m_y\}}.
                                         \label{S}
\ee
Thus, when dealing with continuous variables, we divide the MI either by the sum or by the maximum of the
dimensions. When starting with scalar variables and when $X$ is a cluster variable obtained by joining $m$
elementary variables, then its dimension is just $m_x=m$.

\section{A Phylogenetic Tree for Mammals}

As a first application, we study the mitochondrial DNA of a group of 34 mammals (see Fig.~1).  Exactly the same
data \cite{Genebank} had previously been analyzed in \cite{li1,Reyes00}. This group includes among
others\footnote{opossum (\textit{Didelphis virginiana}), wallaroo (\textit{Macropus robustus}), and platypus
(\textit{Ornithorhyncus anatinus})} some rodents\footnote{rabbit (\textit{Oryctolagus cuniculus}), guinea pig
(\textit{Cavia porcellus}), fat dormouse (\textit{Glis glis}), rat (\textit{Rattus norvegicus}), squirrel
(Scuirus vulgaris), and mouse (\textit{Mus musculus})}, ferungulates\footnote{horse (\textit{Equu caballus}),
donkey (\textit{Equus asinus}), Indian rhinoceros (\textit{Rhinoceros unicornis}), white rhinoceros
(\textit{Ceratotherium simum}), harbor seal (\textit{Phoca vitulina}), grey seal (\textit{Halichoerus grypus}),
cat (\textit{Felis catus}), dog (\textit{Canis familiaris}), fin whale (\textit{Balenoptera physalus}), blue
whale (\textit{Balenoptera musculus}), cow (\textit{Bos taurus}), sheep (\textit{Ovis aries}), pig (\textit{Sus
scrofa}), hippopotamus (\textit{Hippopotamus amphibius}), neotropical fruit bat (\textit{Artibeus jamaicensis}),
African elephant (\textit{Loxodonta africana}), aardvark (\textit{Orycteropus afer}), and armadillo
(\textit{Dasypus novemcintus})}, and primates\footnote{human (\textit{Homo sapiens}), common chimpanzee
(\textit{Pan troglodytes}), pigmy chimpanzee (\textit{Pan paniscus}), gorilla (\textit{Gorilla gorilla}),
orangutan (\textit{Pongo pygmaeus}), gibbon (\textit{Hylobates lar}), and baboon (\textit{Papio hamadryas})}. It
had been chosen in \cite{li1} because of doubts about the relative closeness among these three groups
\cite{cao,Reyes00}.

Obviously, we are here dealing with the algorithmic version of information theory, and informations are
estimated by lossless data compression. For constructing the proximity matrix between individual taxa, we
proceed essentially a in Ref.~\cite{li1}. But in addition to using the special compression program GenCompress
\cite{GenComp}, we also tested several general purpose compression programs such as BWTzip \cite{BWTzip} and the
UNIX tool bzip2.

\begin{figure}
  \begin{center}
   \psfig{file=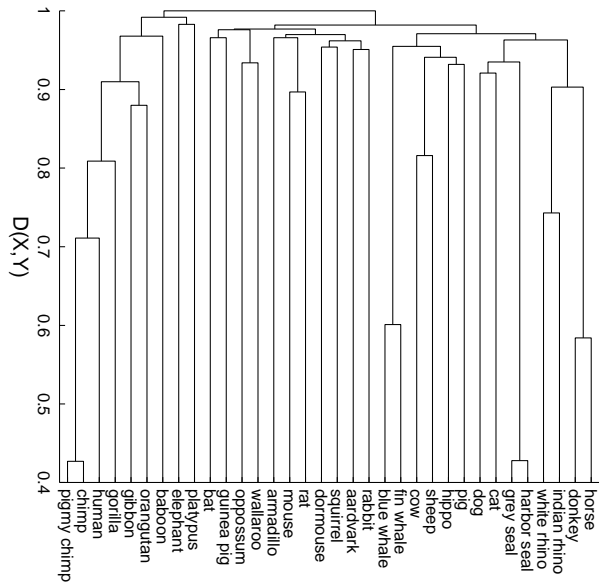,height=80mm,angle=270}
    \caption{Phylogenetic tree for 34 mammals (31 eutherians plus 3 non-placenta mammals).
       In contrast to Fig.~\ref{ClustECG}, the heights of nodes are equal to the distances
       between the joining daughter clusters.}
    \label{phylotree}
\end{center}
\vspace{-8mm}
\end{figure}

In Ref.~\cite{li1}, this proximity matrix was then used as the input to a standard HC algorithm
(neighbour-joining and hypercleaning) to produce an evolutionary tree. It is here where our treatment deviates
crucially. We used the MIC algorithm described in Sec.~1, with distance $D(X,Y)$. The joining of two clusters
(the third step in the MIC algorithm) is obtained by simply concatenating the DNA sequences. There is of course
an arbitrariness in the order of concatenation sequences: $XY$ and $YX$ give in general compressed sequences of
different lengths. But we found this to have negligible effect on the evolutionary tree. The resulting
evolutionary tree obtained with Gencompress is shown in Fig.~\ref{phylotree}.

As shown in Fig.~\ref{phylotree} the overall structure of this tree closely resembles the one shown in
Ref.~\cite{Reyes00}. All primates are correctly clustered and also the relative order of the ferungulates is in
accordance with Ref.~\cite{Reyes00}. On the other hand, there are a number of connections which obviously do not
reflect the true evolutionary tree, see for example the guinea pig with bat and elephant with platypus. But the
latter two, inspite of being joined together, have a very large distance from each other, thus their clustering
just reflects the fact that neither the platypus nor the elephant have other close relatives in the sample. All
in all, however, already the results shown in Fig.~1 capture surprisingly well the overall structure shown in
Ref. \cite{Reyes00}. Dividing MI by the total information is essential for this success. If we had used the
non-normalized $I_{\rm alg}(X,Y)$ itself, the clustering algorithm used in \cite{li1} would not change much,
since all 34 DNA sequences have roughly the same length. But our MIC algorithm would be completely screwed up:
After the first cluster formation, we have DNA sequences of very different lengths, and longer sequences tend
also to have larger MI, even if they are not closely related.

A heuristic reasoning for the use of MIC for the reconstruction of an evolutionary tree might be given as
follows: Suppose that a proximity matrix has been calculated for a set of DNA sequences and the smallest
distance is found for the pair $(X,Y)$. Ideally, one would remove the sequences $X$ and $Y$, replace them by the
sequence of the common ancestor (say $Z$) of the two species, update the proximity matrix to find the smallest
entry in the reduced set of species, and so on. But the DNA sequence of the common ancestor is not available.
One solution might be that one tries to reconstruct it by making some compromise between the sequences $X$ and
$Y$. Instead, we essentially propose to concatenate the sequences $X$ and $Y$. This will of course not lead to a
plausible sequence of the common ancestor, but it will {\it optimally represent the information} about the
common ancestor. During the evolution since the time of the ancestor $Z$, some parts of its genome might have
changed both in $X$ and in $Y$. These parts are of little use in constructing any phylogenetic tree. Other parts
might not have changed in either. They are recognized anyhow by any sensible algorithm. Finally, some parts of
its genome will have mutated significantly in $X$ but not in $Y$, and vice versa. This information is essential
to find the correct way through higher hierarchy levels of the evolutionary tree, and it is preserved in
concatenating.

\section{Clustering of Minimally Dependent Components in an Electrocardiogram}

As our second application we choose a case where Shannon theory is the proper setting. We show in Fig.~2 an ECG
recorded from the abdomen and thorax of a pregnant woman \cite{ECGdata} (8 channels, sampling rate 500 Hz,
5$\,$s total). It is already seen from this graph that there are at least two important components in this ECG:
the heartbeat of the mother, with a frequency of $\approx 3$ beat/s, and the heartbeat of the fetus with roughly
twice this frequency. Both are not synchronized. In addition there is noise from various sources (muscle
activity, measurement noise, etc.). While it is easy to detect anomalies in the mother's ECG from such a
recording, it would be difficult to detect them in the fetal ECG.

As a first approximation we can assume that the total ECG is a linear superposition of several independent
sources (mother, child, noise$_1$, noise$_2$,...). A standard method to disentangle such superpositions is {\it
independent component analysis} (ICA) \cite{ICA}. In the simplest case one has $n$ independent sources
$s_i(t),\; i=1\ldots n$ and $n$ measured channels $x_i(t)$ obtained by instantaneous superpositions with a time
independent non-singular matrix ${\bf A}$,
\be
   x_i(t) = \sum_{j=1}^n A_{ij} s_j(t)\;.
\ee
In this case the sources can be reconstructed by applying the inverse transformation ${\bf W} = {\bf A}^{-1}$
which is obtained by minimizing the (estimated) mutual informations between the transformed components $y_i(t) =
\sum_{j=1}^n W_{ij} x_j(t)$. If some of the sources are Gaussian, this leads to ambiguities \cite{ICA}, but it
gives a unique solution if the sources have more structure.

In reality things are not so simple. For instance, the sources might not be independent, the number of sources
(including noise sources!) might be different from the number of channels, and the mixing might involve delays.
For the present case this implies that the heartbeat of the mother is seen in several reconstructed components
$y_i$, and that the supposedly ``independent" components are not independent at all. In particular, all
components $y_i$ which have large contributions from the mother form a cluster with large intra-cluster MIs and
small inter-cluster MIs. The same is true for the fetal ECG, albeit less pronounced.
It is thus our aim to \\
1) optimally decompose the signals into least dependent components;\\
2) cluster these components hierarchically such that the most dependent ones are
grouped together;\\
3) decide on an optimal level of the hierarchy, such that the clusters make most sense
physiologically;\\
4) project onto these clusters and apply the inverse transformations to obtain cleaned signals for the sources
of interest.

\begin{figure}
  \begin{center}
    \psfig{file=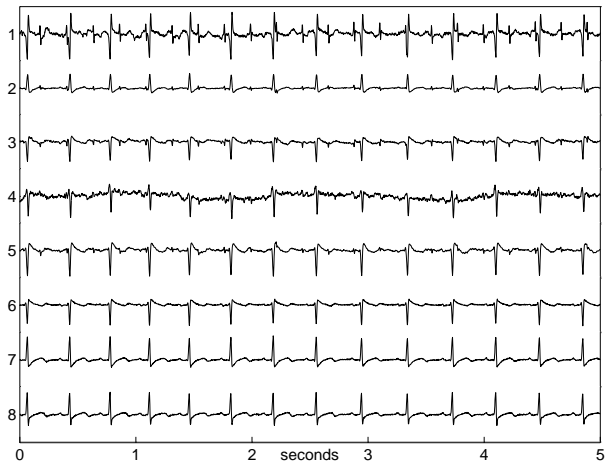,width=80mm}
    \caption{ECG of a pregnant woman.}
    \label{ICAECG0}
\end{center}
\vspace{-8mm}
\end{figure}

\begin{figure}
  \begin{center}
    \psfig{file=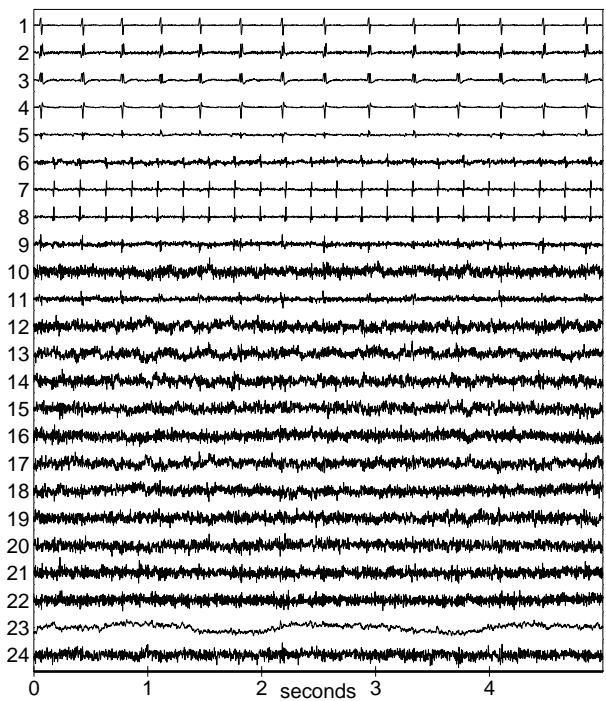,width=80mm}
    \caption{Least dependent components of the ECG shown in Fig.~\ref{ICAECG0}, after increasing
     the number of channels by delay embedding.}
    \label{ICAECG}
\end{center}
\vspace{-8mm}
\end{figure}

Technically we proceeded as follows \cite{Harald}:

Since we expect different delays in the different channels, we first used Takens delay embedding \cite{Takens80}
with time delay 0.002$\,$s and embedding dimension 3, resulting in $24$ channels. We then formed 24 linear
combinations $y_i(t)$ and determined the de-mixing coefficients $W_{ij}$ by minimizing the overall mutual
information between them, using the MI estimator proposed in \cite{mi}. There, two classes of estimators were
introduced, one with square and the other with rectangular neighbourhoods. Within each class, one can use the
number of neighbours, called $k$ in the following, on which the estimate is based. Small values of $k$ lead to a
small bias but to large statistical errors, while the opposite is true for large $k$. But even for very large
$k$ the bias is zero when the true MI is zero, and it is systematically such that absolute values of the MI are
underestimated. Therefore this bias does not affect the determination of the optimal de-mixing matrix. But it
depends on the dimension of the random variables, therefore large values of $k$ are not suitable for the
clustering. We thus proceeded as follows: We first used $k=100$ and square neighbourhoods to obtain the least
dependent components $y_i(t)$, and then used $k=3$ with rectangular neighbourhoods for the clustering. The
resulting least dependent components are shown in Fig.~\ref{ICAECG}. They are sorted such that the first
components (1 - 5) are dominated by the mother's ECG, while the next three contain large contributions from the
fetus. The rest contains mostly noise, although some seem to be still mixed.

\begin{figure}
  \begin{center}
    \psfig{file=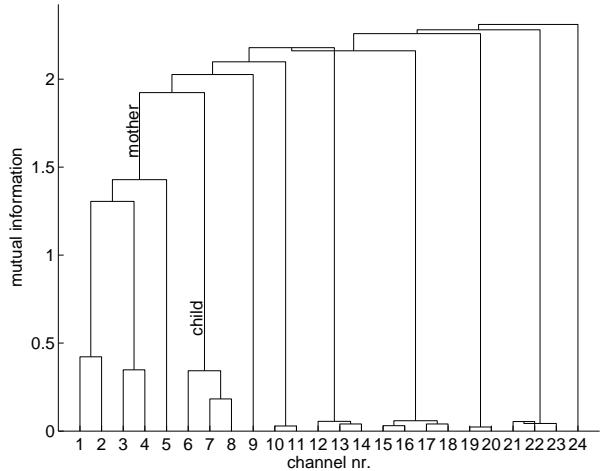,width=80mm,angle=0}
    \caption{Dendrogram for least dependent components. The height where the two branches of
    a cluster join corresponds to the MI of the cluster.}
    \label{ClustECG}
\end{center}
\vspace{-8mm}
\end{figure}

These results obtained by visual inspection are fully supported by the cluster analysis. The dendrogram is shown
in Fig.~\ref{ClustECG}. In constructing it we used $S(X,Y)$ (Eq.(\ref{S})) as similarity measure to find the
correct topology. Again we would have obtained much worse results if we had not normalized it by dividing MI by
$m_X+m_Y$. In plotting the actual dendrogram, however, we used the MI of the cluster to determine the height at
which the two daughters join. The MI of the first five channels, e.g., is $\approx 1.43$, while that of channels
6 to 8 is $\approx 0.34$. For any two clusters (tuples) $X=X_1\ldots X_n$ and $Y=Y_1\ldots Y_m$ one has $I(X,Y)
\geq I(X)+I(Y)$. This guarantees, if the MI is estimated correctly, that the tree is drawn properly. The two
slight glitches (when clusters (1--14) and (15--18) join, and when (21--22) is joined with 23) result from small
errors in estimating MI. They do in no way effect our conclusions.

\begin{figure}
  \begin{center}
    \psfig{file=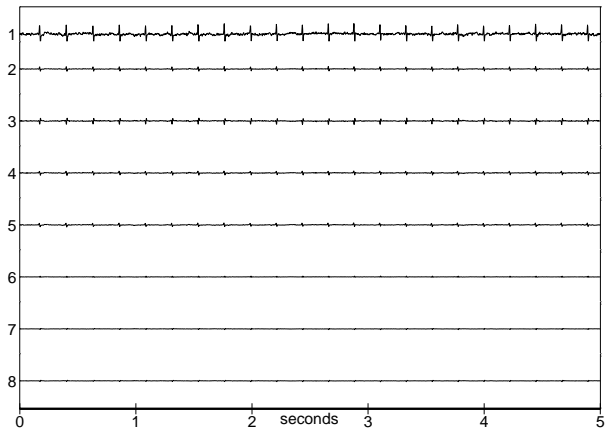,width=80mm}
    \caption{Original ECG where all contributions except those of the child cluster have
     been removed.}
    \label{reconstruct}
\vspace{-8mm}
\end{center}
\end{figure}

In Fig.~\ref{ClustECG} one can clearly see two big clusters corresponding to the mother and to the child. There
are also some small clusters which should be considered as noise. For reconstructing the mother and child
contributions to Fig.~\ref{ICAECG0}, we have to decide on one specific clustering from the entire hierarchy. We
decided to make the cut such that mother and child are separated. The resulting clusters are indicated in
Fig.~\ref{ClustECG} and were already anticipated in sorting the channels. Reconstructing the original ECG from
the child components only, we obtain Fig.~\ref{reconstruct}.

\section{Conclusions}

We have shown that MI can not only be used as a proximity measure in clustering, but that it also suggests a
conceptually very simple and natural hierarchical clustering algorithm. We do not claim that this algorithm,
called {\it mutual information clustering} (MIC), is always superior to other algorithms. Indeed, MI is in
general not easy to estimate. Obviously, when only crude estimates are possible, also MIC will not give optimal
results. But as MI estimates are becoming better, also the results of MIC should improve. The present paper was
partly triggered by the development of a new class of MI estimators for continuous random variables which have
very small bias and also rather small variances \cite{mi}.

We have illustrated our method with two applications, one from genetics and one from cardiology. For neither
application MIC might give the very best clustering, but it seems promising that one common method gives decent
results for both, although they are very different.

The results of MIC should improve, if more data become available. This is trivial, if we mean by that longer
time sequences in the application to ECG, and longer parts of the genome in the application of Sec.3. It is less
trivial that we expect MIC to make fewer mistakes in a phylogenetic tree, when more species are included. The
reason is that close-by species will be correctly joined anyhow, and families -- which now are represented only
by single species and thus are poorly characterized -- will be much better described by the concatenated genomes
if more species are included.

There are two versions of information theory, algorithmic and probabilistic, and therefore there are also two
variants of MI and of MIC. We discussed in detail one application of each, and showed that indeed common
concepts were involved in both. In particular it was crucial to normalize MI properly, so that it is essentially
the {\it relative} MI which is used as proximity measure. For conventional clustering algorithms using
algorithmic MI as proximity measure this had already been stressed in \cite{li1,li2}, but it is even more
important in MIC, both in the algorithmic and in the probabilistic versions.

In the probabilistic version, one studies the clustering of probability distributions. But usually distributions
are not provided as such, but are given implicitly by finite random samples drawn (more or less) independently
from them. On the other hand, the full power of algorithmic information theory is only reached for infinitely
long sequences, and in this limit any individual sequence defines a sequence of probability measures on finite
subsequences. Thus the strict distinction between the two theories is somewhat blurred in practice.
Nevertheless, one should not confuse the similarity between two sequences (two English books, say) and that
between their subsequence statistics. Two sequences are maximally different if they are completely random, but
their statistics for short subsequences is then identical (all subsequences appear in both with equal
probabilities). Thus one should always be aware of what similarities or independencies one is looking for. The
fact that MI can be used in similar ways for all these problems is not trivial.

We would like to thank Arndt von Haesseler, Walter Nadler and Volker Roth for many useful discussions.

\bibliographystyle{apalike}
\bibliography{lit}

\begin{thebibliography}{}

\bibitem[Gen, a]{Genebank}
\textsf{http://www.ncbi.nlm.nih.gov}.

\bibitem[Gen, b]{GenComp}
\textsf{http://www.cs.ucsb.edu/~mli/Bioinf/software}.

\bibitem[BWT, ]{BWTzip}
\textsf{http://stl.caltech.edu/bwtzip.shtml}.

\bibitem[Cao et~al., 1998]{cao}
Cao, Y., Janke, A., Waddell, P., Westerman, M., Takenaka, O., Murata, S.,
  Okada, N., P{\"a}{\"a}bo, S., and Hasegawa, M. (1998).
\newblock Conflict among individual mitochondrial proteins in resolving the
  phylogeny of eutherian orders.
\newblock {\em J. Molec. Evol.}, 47(3):307--322.

\bibitem[Cover and Thomas, 1991]{cover-thomas}
Cover, T.~M. and Thomas, J.~A. (1991).
\newblock {\em Elements of Information Theory}.
\newblock John Wiley and Sons, New York.

\bibitem[{De Moor}, 1997]{ECGdata}
{De Moor}, B. L.~R. (1997).
\newblock Daisy: Database for the identification of systems.
\newblock \textsf{www.esat.kuleuven.ac.be/sista/daisy}.

\bibitem[Hyv{\"a}rinen et~al., 2001]{ICA}
Hyv{\"a}rinen, A., Karhunen, J., and Oja, E. (2001).
\newblock {\em Independent component analysis}.
\newblock John Wiley and Sons, New York.

\bibitem[Jain and Dubes, 1988]{jain-dubes}
Jain, A.~K. and Dubes, R.~C. (1988).
\newblock {\em Algorithms for Clustering Data}.
\newblock Prentice Hall, Englewood Cliffs, NJ.

\bibitem[Kraskov et~al., 2003]{mi}
Kraskov, A., St{\"o}gbauer, H., and Grassberger, P. (2003).
\newblock Estimating mutual information.
\newblock E-print, ar\textsc{X}iv.org/cond-mat/0305641.

\bibitem[Li et~al., 2001]{li1}
Li, M., Badger, J.~H., Chen, X., Kwong, S., Kearney, P., and Zhang, H. (2001).
\newblock An information-based sequence distance and its application to whole
  mitochondrial genome phylogeny.
\newblock {\em Bioinformatics}, 17(2):149--154.

\bibitem[Li et~al., 2002]{li2}
Li, M., Chen, X., Li, X., Ma, B., and Vitanyi, P. (2002).
\newblock The similarity metric.
\newblock E-print, ar\textsc{x}iv.org/cs.CC/0111054.

\bibitem[Li and Vitanyi, 1997]{li-vi}
Li, M. and Vitanyi, P. (1997).
\newblock {\em An introduction to Kolmogorov complexity and its applications}.
\newblock Springer, New York.

\bibitem[Reyes et~al., 2000]{Reyes00}
Reyes, A., Gissi, C., Pesole, G., Catzeflis, F.~M., and Saccone, C. (2000).
\newblock Where do rodents fit? \textsc{E}vidence from the complete
  mitochondrial genome of sciurus vulgaris.
\newblock {\em Mol. Biol. Evol.}, 17:979--983.

\bibitem[St{\"o}gbauer et~al., 2004]{Harald}
St{\"o}gbauer, H., Kraskov, A., and Grassberger, P. (2004).
\newblock Potential of mutual information in application to \textsc{ICA}.

\bibitem[Takens, 1980]{Takens80}
Takens, F. (1980).
\newblock Detecting strange attractors in turbulence.
\newblock In Rand, D.~A. and Young, L.~S., editors, {\em Dynamical Systems and
  Turbulence}, volume 898 of {\em Lecture Notes in Mathematics}, page 366.
  Springer-Verlag, Berlin.

\bibitem[Tishby et~al., 1999]{bottleneck}
Tishby, N., Pereira, F., and Bialek, W. (1999).
\newblock The information bottleneck method.
\newblock In {\em 37-th Annual Allerton Conference on Communication, Control
  and Computing}, page 368.

\end{thebibliography}

\end{document}